\documentclass[twocolumn]{webofc}

\usepackage[varg]{txfonts}   
\usepackage{siunitx}
\usepackage{todonotes}
\begin{document}
\title{nEDM experiment at PSI: data-taking strategy and sensitivity of the dataset}

\author{\lastname{C.~Abel} \inst{1} \and \lastname{N.~J.~Ayres} \inst{1} \and \lastname{G.~Ban} \inst{2} \and \lastname{G.~Bison} \inst{3} \and \lastname{K.~Bodek} \inst{4} \and \lastname{V.~Bondar} \inst{5} \and \lastname{E.~Chanel} \inst{6} \and \lastname{P.-J.~Chiu} \inst{3,7} \and \lastname{M.~Daum} \inst{3} \and \lastname{S.~Emmenegger} \inst{7} \and  \lastname{L.~Ferraris-Bouchez} \inst{8} \fnsep\thanks{\email{ferraris@lpsc.in2p3.fr}} \and \lastname{P.~Flaux} \inst{2} \and \lastname{W.~C.~Griffith} \inst{1} \and \lastname{P.~G.~Harris} \inst{1} \and \lastname{N.~Hild} \inst{3} \and \lastname{Y.~Kermaidic} \inst{8} \fnsep\thanks{Present address: Max Planck Institute für Kernphysik, Heidelberg, Germany} \and \lastname{K.~Kirch} \inst{3,7} \and \lastname{P.~A.~Koss} \inst{5} \and \lastname{J.~Krempel} \inst{7} \and \lastname{B.~Lauss} \inst{3} \and \lastname{T.~Lefort} \inst{2} \and \lastname{Y.~Lemiere} \inst{2} \and \lastname{A.~Leredde} \inst{8} \and \lastname{P.~Mohanmurthy} \inst{3,7} \and \lastname{M.~Musgrave} \inst{1} \fnsep\thanks{Present address: Massachusetts Institute of Technology, Cambridge, MA, USA} \and \lastname{O.~Naviliat-Cuncic} \inst{2} \fnsep\thanks{Present address:  Michigan State University, East-Lansing, USA} \and \lastname{D.~Pais} \inst{3,7} \and \lastname{F.~M.~Piegsa} \inst{6} \and \lastname{G.~Pignol} \inst{8} \and \lastname{M.~Rawlik} \inst{7} \and \lastname{D.~Rebreyend} \inst{8} \and \lastname{D.~Ries} \inst{9} \and \lastname{S.~Roccia}  \inst{10} \fnsep\thanks{\email{roccia@csnsm.in2p3.fr}, present address: Institut Laue-Langevin, Grenoble, France} \and \lastname{D.~Rozpedzik} \inst{4} \and \lastname{P.~Schmidt-Wellenburg} \inst{3}  \fnsep\thanks{\email{philipp.schmidt-wellenburg@psi.ch}} \and \lastname{A.~Schnabel} \inst{11}  \and \lastname{N.~Severijns} \inst{5}  \and \lastname{J.~Thorne} \inst{1} \and \lastname{R.~Virot} \inst{8} \and \lastname{J.~Voigt} \inst{11} \and \lastname{A.~Weis} \inst{12} \and \lastname{E.~Wursten} \inst{5} \and \lastname{J.~Zejma} \inst{4} \and \lastname{G.~Zsigmond} \inst{3}}	

\institute{Department of Physics and Astronomy, University of Sussex, Brighton, United Kingdom 
\and 
Normandie Univ, ENSICAEN, UNICAEN, CNRS/IN2P3, LPC Caen, Caen, France 
\and
Paul Scherrer Institute (PSI), Villigen, Switzerland 
\and
Marian Smoluchowski Institute of Physics, Jagiellonian University, Cracow, Poland 
\and
Instituut voor Kern-- en Stralingsfysica, Katholieke~Universiteit~Leuven, Leuven, Belgium 
\and
Laboratory for High Energy Physics, Albert Einstein Center for Fundamental Physics, University of Bern, Bern, Switzerland 
\and
ETH Z\"urich, Institute for Particle Physics and Astrophysics, Z\"urich, Switzerland 
\and
Univ. Grenoble Alpes, CNRS, Grenoble INP, LPSC-IN2P3, Grenoble, France 
\and
Institute of Nuclear Chemistry, Johannes Gutenberg University, Mainz, Germany 
\and 
CSNSM, Universit\'e Paris Sud, CNRS/IN2P3, Universit\'e Paris Saclay, Orsay-Campus, France 
\and 
Physikalisch Technische Bundesanstalt, Berlin, Germany 
\and
University of Fribourg, Fribourg, Switzerland 
}

\abstract{%
  We report on the strategy used to optimize the sensitivity of our search for a neutron electric dipole moment at the Paul Scherrer Institute. Measurements were made upon ultracold neutrons stored within a single chamber at the heart of our apparatus. A mercury cohabiting magnetometer together with an array of cesium magnetometers were used to monitor the magnetic field, which was controlled and shaped by a series of precision field coils. In addition to details of the setup itself, we describe the chosen path to realize an appropriate balance between achieving the highest statistical sensitivity alongside the necessary control on systematic effects.  The resulting irreducible sensitivity is better than $1\!\times\!10^{-26}e{\rm cm}$. This contribution summarizes in a single coherent picture the results of the most recent publications of the collaboration.    
}
\maketitle
\section{Introduction}
\label{intro}
Establishing an appropriate strategy for taking data in a  neutron electric dipole moment (nEDM) measurement requires facing two challenges: accumulating the statistics in a very efficient way (such experiments tend to be ultimately limited by statistics) while keeping the systematic effects under control continuously. This is well illustrated by the best upper limit to date \cite{Pendlebury2015}: 
\begin{equation}
d_n=(-0.21 \pm 1.53~\textmd{(stat)} \pm 0.99~\textmd{(syst)})\times 10^{-26}~e\textmd{cm},
\end{equation}
where the contributions from systematic effects are clearly non-negligible. To overcome these difficulties we have combined a number of offline measurements with an optimized data-taking sequence in which a large number of parameters were changed in a systematic manner. This strategy has enabled us to control systematic effects to an unprecedented level, in particular with respect to the inhomogeneities of the magnetic field, and it thus paves the way towards the new generation of experiments currently being established \cite{Ito2007, Altarev2012, Picker2017, Serebrov2017, Abel2018Proc}.

\section{The Ramsey spectrometer at the Paul Scherrer Institute}
\label{spectro}
The largely refitted and upgraded nEDM spectrometer installed at the Paul Scherrer Institute~(PSI) from 2009 to 2017 was in part originally operated at the Institut Laue-Langevin (ILL) from 1996 onwards, during which time it successfully pushed down the upper limit on the neutron electric dipole moment \cite{Baker2006, Harris1999}; indeed, some components -- in particular, the magnetic shields -- had also been used in a previous world-limit measurement \cite{Smith1990}. The current nEDM upper limit \cite{Pendlebury2015} is based on a reanalysis of the 1998-2002 dataset from ILL.  The spectrometer uses a single \SI{20}{\liter} cylindrical ultracold-neutron (UCN) storage chamber mounted in a four layer mu-metal shield with a quasi-static shielding factor of up to 10\,000. The storage chamber (Ramsey precession cell) installed at PSI consisted of  upper and lower electrodes made of DLC-coated\,\cite{Atchison2006PRC} aluminum plates, a deuterated polystyrene coated Rexolite cylinder\,\cite{Bodek2008}, and deuterated polyethylene coated optical quartz windows\,\cite{Bodek2008}.
It was designed to store both polarized ultracold neutrons and polarized mercury atoms ($\rm ^{199}Hg$), the latter being used as a cohabiting magnetometer.
The two species precessed in a $B_0=\SI{1}{\micro\tesla}$ highly homogeneous vertical magnetic field, and were also exposed to an $E=\SI{11}{kV/cm}$ vertical electric field. Both fields were regularly reversed; a process that took a few minutes for the electric field and a few hours for the magnetic field. 

\begin{figure}
	\centering
	\hspace*{-0.5cm}\includegraphics[width=1.1\columnwidth]{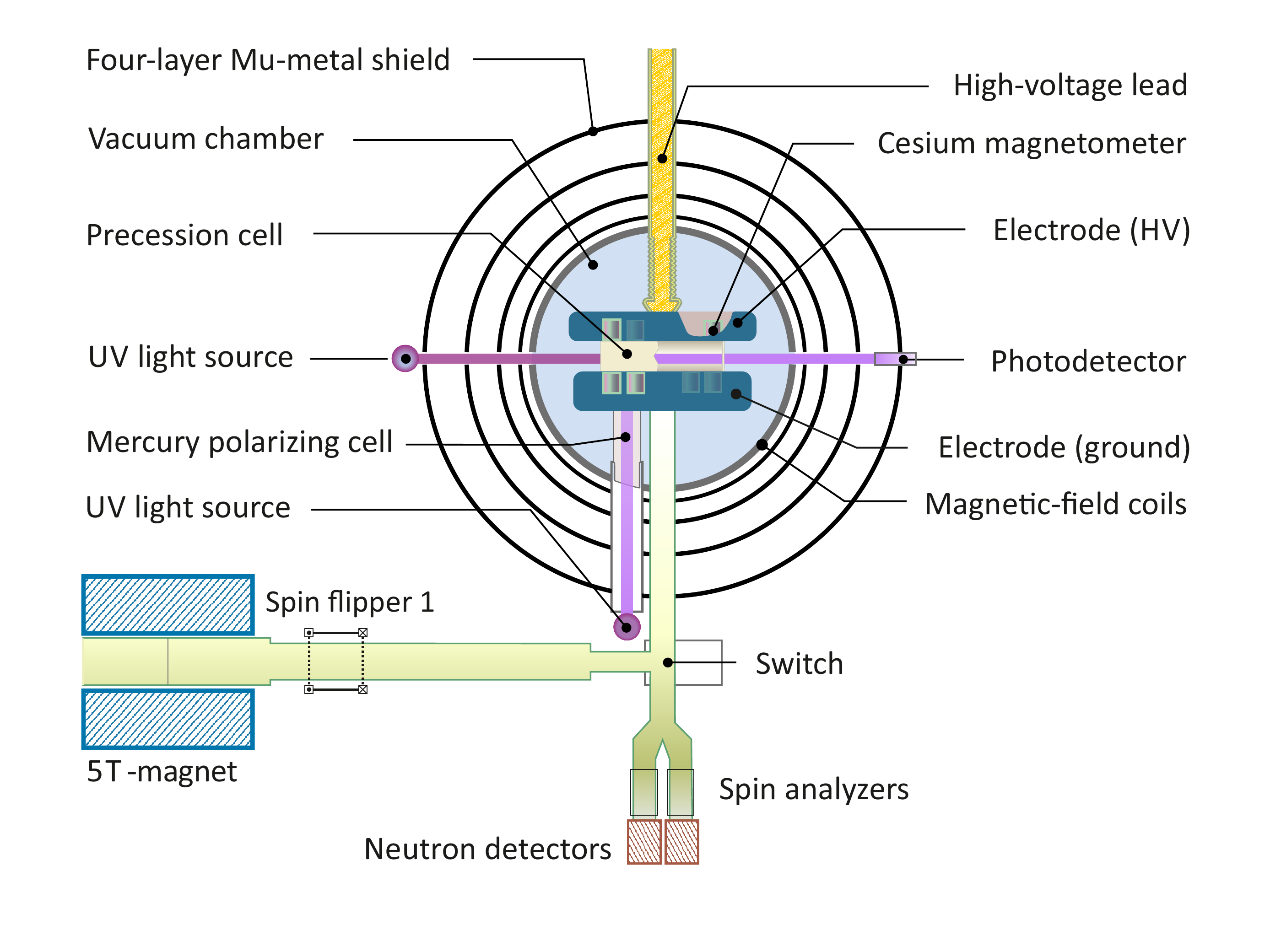}%
	\caption{Sketch of the spectrometer used to search for an electric-field induced shift in the magnetic resonance frequency of polarized UCN exposed to an electric field of strength $E=\SI{11}{kV/cm}$ and a magnetic field of $B_0 =\SI{1.035}{\micro\tesla}$.}%
	\label{fig:apparatus}%
\end{figure}

Figure\,\ref{fig:apparatus} shows a sketch of the apparatus. The full dataset consists of more than 50\,000 single measurements, a.k.a.\ cycles, which were grouped in sets having the same magnetic-field configuration. Cycles were repeated every 300~s, and each followed the same sequence of events.
UCN from the source\,\cite{Lauss2014} were first guided through a \SI{5}{\tesla} superconducting magnet that fully polarized the neutrons by reflecting all spin-up neutrons back towards the source. Immediately following the superconducting magnet an adiabatic fast-passage spin-flipper (referred to as SF1) was used to choose the initial neutron spin state.
The neutrons were then deflected by a switch into the precession chamber. After a filling period of 28 s, the entrance door to the storage chamber was closed. That duration was determined through measurements to optimize the product $N\alpha_0^2$ between the number of neutrons, $N$, and the square of the initial polarization $\alpha_0$. During the next \SI{2}{s} the polarized mercury atoms were admitted into the precession chamber.

The precession was initiated by two successive \SI{2}{s} long $\pi/2$ pulses matching the mercury ($\sim 8\,$Hz) and (approximately) the neutron ($\sim 30\,$Hz) Larmor frequencies. For the neutrons this was the first step of Ramsey's technique of separated oscillating fields. Both the neutrons and the mercury atoms then precessed freely for 180~s. The precession frequency of the mercury atoms was continuously probed by an ultraviolet light beam traversing the storage chamber. The neutron precession frequency was measured by applying at the end of the precession period a second $\pi/2$ pulse in phase with the first, and measuring the resulting distribution of neutrons in each spin state. To that end, neutrons were guided to the detection system \cite{Afach2015} which simultaneously counted the two spin polarities in each of two different detectors \cite{Ban2016}. Neutrons entering the detection system had to pass through a magnetization-saturated polarized iron foil before reaching either of the two detectors (named A and B). This foil reflected spin-up neutrons with an efficiency of $90(4)\%$. Each arm of the detector incorporated a spin flipper, with a measured flip efficiency of $97(1)\%$, above the iron foil.  By switching on either of the two spin flippers   (referred to as SF2$_{\rm A/B}$), the detector in question could be made to count spin-up instead of spin-down neutrons. 
This detection system allowed the interchange of detectors detecting a given spin state.
We used this feature associated with a periodic switching of the SF1 spin flipper (efficiency $95(3)$\%) to avoid potential systematic effects that might arise from an asymmetry in the detection efficiency.

The high voltage for the electric field was delivered by a true bipolar \SI{200}{kV} voltage~(HV) supply\footnote{HCB~40M - 200\,000, FuG Elektronik GmbH, https://www.fug-elektronik.de/en/}, \SI{200}{\micro\ampere}, with a maximum ramp speed of \SI{1}{kV/s}.

The magnetic field was made uniform through the use of 36 trim coils, of which 30 were directly wound onto the aluminum vacuum tank as saddle coils or cylindrical coils and another six solenoids were wound onto cylinders (UCN guide, HV feedthrough, and mercury polarization chamber) that penetrated the mu-metal shield through vertical holes. An array of 16 optically pumped cesium magnetometers\,\cite{ElisePaper} mounted above and below the electrodes was used to monitor the magnetic field.

The entire experimental installation was mounted inside a thermally insulated cabin, divided into a zone around the mu-metal shield that was stabilized to better than \SI{0.1}{\kelvin} (per 24h) and a control room that accommodated all of the electronics and that was stabilized to \SI{1}{\kelvin} (per 24h). 
Six large rectangular coils (one pair per spatial direction) were attached to the cabin. 
Using a set of 30 fluxgates in the vicinity of the mu-metal shield, the coil currents were actively stabilized  \cite{Afach2014JAP} to suppress the environmental magnetic field and to compensate for field changes due to magnetic activity in the laboratory in particular to a large superconducting solenoid that was used for quench tests of superconductor prototypes (see references in \cite{Abel2018Proc}).

	
\pagebreak
\section{Data-taking strategy}
\label{strategy}

For the purpose of minimizing unintentional bias during analysis the vast majority of the data were blinded. After an initial period of one month in 2015 taking non-blinded data and testing the blinding algorithm, described in more detail in Ref.\,\cite{Krempel2018}, an unknown, artificial neutron EDM value (randomly generated during compilation of the code) was injected by marginally changing the counting statistics of the detectors A and B. The original data were encrypted and saved on a password protected server, while the two analysis groups obtained sets of data to which in a first step a common offset and in a second step a group-specific offset was applied. This guaranteed that the central values differed between the two groups, and then allowed for a relative unblinding by removal of the second individual blinding offsets so that the results of the two groups could be compared while still preserving the common offset. Once both groups have finished their analyses and it has been established that their results agree after removal of the secondary blinding, both groups' analyses will be run on the original unblinded data in order to provide the final result.


\subsection{The crossing-point analysis} \label{CrossPointAna}
In an apparatus with a single storage chamber, the neutron electric dipole moment is measured by searching for a change in the neutron precession frequency $f_{ \rm {n}}$ due to a reversal in the relative directions of electric and magnetic fields from parallel to antiparallel: 
\begin{equation}
d_n=  h \frac{ f_{ \rm {n}, \uparrow \downarrow} - f_{ \rm {n}, \uparrow \uparrow}}{4 E}.
\end{equation}
The mercury co-magnetometer is, in this geometry, primarily used to compensate for the unavoidable fluctuations of the magnetic field which would otherwise be the main limiting factor for the sensitivity. One can form a quantity $\mathcal{R}$ sensitive to such relative changes in frequency that, to first order, does not depend upon the magnetic field: 
\begin{equation}
 \mathcal{R} = \frac{f_{\rm n}}{f_{\rm Hg}} = \left |{\frac{\gamma_{\rm n}}{\gamma_{\rm Hg}} }\right | \left( 1 + \delta_{\rm grav}  + \delta_{\rm T} +\delta_{\rm nEDM}+ \delta_{\rm other} \right).
\label{Reqt}
\end{equation}
It is clear from this expression that the ratio of the neutron precession frequency  $f_{\rm n}$ to the mercury precession frequency $f_{\rm Hg}$ is not precisely equal to the ratio of gyromagnetic ratios $\gamma$. The correction factors are listed in decreasing order. The first and largest correction, the gravitational shift, is due to the difference between the center of mass of the (thermal) mercury atoms, located at the center of the precession chamber, and that of the (ultracold) neutrons $\langle z \rangle$, which is a little lower:
\begin{equation}
\delta_{\rm grav} =\pm \frac{G_{\rm grav} \langle z \rangle}{|B_0|},
\end{equation}
where the $\pm$ sign refers to the direction of the magnetic field. Its amplitude depends on $G_{\rm grav}$, the effective vertical gradient of the magnetic field. To first order $G_{\rm grav}~\approx~\partial B_z / \partial z$ but,  as discussed in Ref.\,\cite{Abel2018}, higher order terms cannot be neglected. The second correction term $\delta_{\rm T}$ depends on the residual transverse component $B_{\rm T}$ of the magnetic field. It arises due to the different behavior of neutrons and mercury atoms which is adiabatic for neutrons and non-adiabatic for mercury atoms within the magnetic field:
\begin{equation}
\delta_{\rm T} = \frac{\langle B_{\rm T}^2 \rangle}{2 B_0^2}. 
\label{deltaT}
\end{equation}
The last term, $\delta_{\rm other}$, includes smaller corrections such as the rotation of the  Earth and the frequency shift inherent to the use of light to probe the mercury precession frequency. 

It is now well known that the mercury co-magnetometer is subject to an electric-field-dependent frequency shift arising from a combined effect of magnetic-field inhomogeneities and of the motional magnetic field $\bf{B_{\rm m}}=\bf{E}\times\bf{v}/c^2$ \cite{Pendlebury2004}. This shift represents the main systematic effect of this experiment. 
It can be exactly calculated in the non-adiabatic regime, where the Larmor frequency is much slower than the wall collision rate, as is the case for mercury atoms, provided that the inhomogeneities of the magnetic field are known\,\cite{Pignol2012}.
When correcting the time variations of the magnetic field using the mercury co-magnetometer, this $E$-dependent mercury frequency shift introduces a false neutron-EDM-mimicking signal of $d^{\rm false}_{\rm n \leftarrow Hg} = \left|\frac{\gamma_{\rm n}}{\gamma_{\rm Hg}}\right| d^{\rm false}_{\rm Hg}$. 
As an example, the false neutron EDM owing to a shift of the mercury precession frequency is calculated up to cubic order for a cylindrical precession chamber of height $H$ and diameter $D$ \cite{Abel2018}. Interestingly, this exact calculation can be divided into one term linear in $G_{\rm grav}$ and a higher-order term (see~\cite{Abel2018} for the exact definition of the cubic term $G_{3,0}$):
\begin{equation}
d_{\rm n}^{\rm meas} = d_{\rm n}^{\rm true} + \frac{\hbar \left|\gamma_{\rm n} \gamma_{\rm Hg}\right|}{32 c^2} D^2 
\left[G_{\rm grav} + G_{3,0} \left( \frac{D^2}{16} + \frac{H^2}{10} \right) \right]. 
\label{FalseEDMGeo}
\end{equation}

By taking advantage of the fact that equations \eqref{Reqt} and \eqref{FalseEDMGeo} show both a linear dependence upon the same gradient $G_{\rm grav}$, one can measure the neutron EDM by setting different values for $G_{\rm grav}$ and obtaining a curve of $d_{\rm n}^{\rm meas}$ versus  $\mathcal{R}$.
Provided that all of the correction factors ($G_{3,0}$, $\delta_{\rm T}$, $\delta_{\rm other}$) are compensated for, the point where $G_{\rm grav}=0$ and thus $d_{\rm n}^{\rm meas} = d_{ \rm n}^{\rm true}$ lies at the crossing point of the two curves $d_{\rm n}^{\rm meas}$ versus $\mathcal{R}$ for the two $B_0$ directions.  Note that $\delta_{\rm grav}$ depends on the sign of $B_0$. This strategy is a revised and extended version of the one pioneered in Ref.\,\cite{Baker2006}.    

There are two reasons why the correction using the so-called crossing point could be more complicated than the simple linear extrapolation described above. On the one hand, local magnetic dipoles create a false EDM larger than the one predicted by equation\,\eqref{FalseEDMGeo}~\cite{Harris2006,Pignol2012}. On the other hand, the gravitationally induced vertical striation of ultracold neutrons\,\cite{Harris2014,Afach2015PRL} in combination with a vertical magnetic-field gradient induces a nonlinear dependence between $ \mathcal{R}$ and $G_{\rm grav}$ which can shift the crossing point away from the point where $d_{\rm n}^{\rm meas} = d_{\rm n}^{\rm true}$. Indeed, this effect triggered the reanalysis of the best limit to date \cite{Pendlebury2015}. Furthermore, this effect depolarizes neutrons and thus has a direct impact upon the achievable sensitivity, as discussed in section~\ref{sensitivity}.    

A well optimized data-taking sequence combined with the fine tuning of adjustable parameters and complemented by a series of auxiliary measurements has permitted us to keep these two effects as well as the other correction factors in equations \eqref{Reqt} and \eqref{FalseEDMGeo} sufficiently well under control to achieve the world's highest sensitivity nEDM search to date. 

\subsection{Auxiliary measurements}

\subsubsection*{Magnetic field maps} 
\begin{figure*}[ttt]
\centering
\includegraphics[width=16cm,clip]{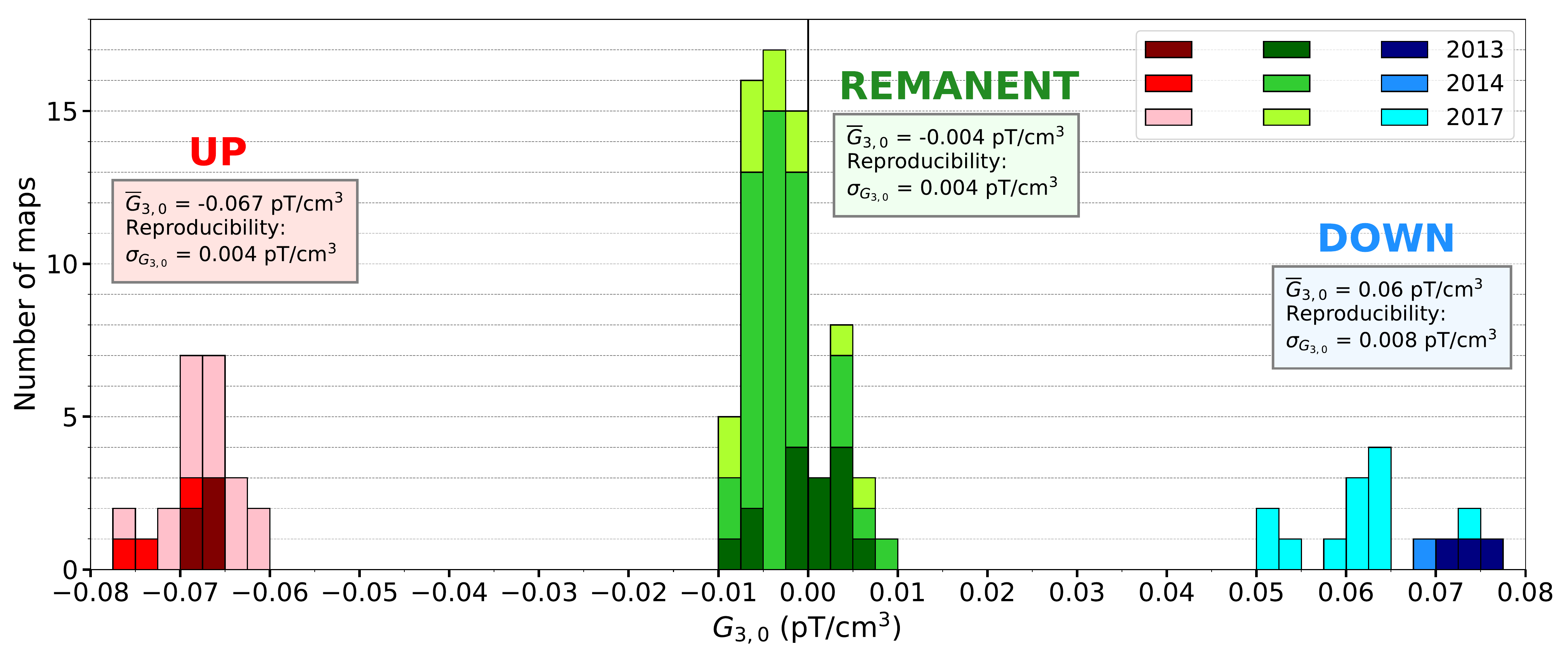}
\caption{Reproducibility study on the extraction of the $G_{3,0}$ cubic term over four years. This study was done for the main coil creating a magnetic field pointing upwards (downwards) as shown in red (blue). Green shows the $G_{3,0}$ cubic terms from maps without any current in the main coil.}
\label{MapCampaign}       
\end{figure*}

In addition to the very accurate measurement by the mercury magnetometer of the average magnetic field within the storage volume, additional information about the profile of the magnetic field, such as $\partial B_z/\partial z$, was provided by the cesium-magnetometer array. 
In order to access higher order inhomogeneities in the magnetic field, offline field maps were generated using three-axis fluxgate magnetometers mounted on an automated non-magnetic rotational/translational support~\cite{FieldMapPaper}. Maps of all of the 36 magnetic fields generated by trim coils, in addition to numerous maps of the field of the main coil (for reproducibility studies), were recorded over periods of several weeks.
Each single field-map measurement typically took about five hours. 
These studies required the dismounting of the precession chamber, the mercury polarization chamber, and a large part of the vacuum system.
We conducted three such campaigns in 2013, in 2014 before starting nEDM data taking, and in 2017 after the end of data taking, always during the annual maintenance period of the proton accelerator. In total we took about 300 maps, including the 105 maps used for reproducibility studies as summarized in Fig.\,\ref{MapCampaign}. 

The first two field-mapping campaigns were used to study the long-term stability and reproducibility of the field maps, and also to inform the tuning of the spectrometer for data taking. The field maps taken in 2014 and 2017 were used to infer the fields generated by each coil during data collection. Additionally, in 2017, we took a map of each magnetic-field configuration that had been used during data taking. In the analysis we use these maps to extract higher order inhomogeneities in the magnetic field such as the transverse component of equation \eqref{deltaT} and the cubic term in equation \eqref{FalseEDMGeo}. For the $G_{3,0}$ cubic term, the limiting factor is the reproducibility of the field, which was found to be better than 0.008~pT/cm$^3$ over five years. The knowledge of the transverse component $B_{\rm T}$ is similarly limited by the reproducibility at the level of 0.4~nT$^2$. 

\subsubsection*{Magnetic scans of experimental components}
We pursued a similar strategy to constrain the presence of local dipoles. All components close to the precession chamber were scanned in the large magnetically shielded room BMSR-2~\cite{PTBpaper} at PTB-Berlin using a SQUID magnetometer. These items ranged in size from the large electrodes of diameter \SI{47}{cm} to millimeter sized screws and bolts. We scanned each item twice: once before data-taking to ensure that there was no significant magnetic contamination at the start, and once after data-taking to ensure that no magnetic contamination had been introduced in the interim. As a result we produced a catalog of upper limits of all dipoles found with their strength and location.  


\subsection{Ramp-up activities}

In each of the years 2015, 2016 and 2017, data-taking started with a few days dedicated to the test and characterization of the spectrometer. It included the measurement of the efficiency of the spin flippers, the optimization of the filling time and of the switch settings, and the measurement of the background in the ultracold neutron detectors. In addition, the quality of the inner surface of the precession chamber was tracked via measurements of the storage times $T_{0,f}$ and $T_{0,s}$ for the fast and slow component respectively and of the longitudinal depolarization time $T_{1}$. Table~\ref{tab-time} shows the evolution of those time constants. One can notice the improvement in $T_{0,s}$ in particular between 2015 and 2016, which is a consequence of the O$_2$ discharge cleaning procedures that we carried out and the improving vacuum conditions over this period. Indeed, the nEDM spectrometer was kept closed continuously throughout the 2015-2016 period. The deterioration in the 2017 value is explained by the necessity to reopen the system in early 2017 to perform auxiliary measurements. 

\begin{table}[ht]
\centering
\caption{Summary of characteristic time constant during the data taking at PSI. }
\label{tab-time}       
\begin{tabular}{l|c|c|c}
 & 2015 & 2016 & 2017 \\
\hline
\hline
$T_{0,f}$~(s) & 75(11)      & 72(2)     & 40(4) \\
$T_{0,s}$~(s) & 271(16)     & 320(7)    & 252(4)\\
$T_{1}$~(s)   & 6000(1300)  & 4200(200) & 5700(500)\\
\end{tabular}
\end{table}

Finally, the annual start-up procedure also included a spin-echo study. As described in \cite{Afach2015PRL}, the spin-echo technique can be used to measure the spectrum of the ultracold neutrons by applying relatively large vertical magnetic gradients within the range $\pm 50$~pT/cm.


\subsection{nEDM sequence}
During the 253~days of data taking, the magnetic field was changed 19 times. The proton accelerator maintenance period  (three days without protons every four weeks) defined a natural time scale to record a full set of measurements with a given magnetic-field configuration. In order to establish a new magnetic configuration the main coil was first of all powered appropriately for the given direction of the magnetic field, and the shield was idealized (i.e.\ degaussed with the magnetic field on; also known as equilibration). After waiting for at least 30 minutes for further relaxation of the mu-metal shield, the magnetic field was stable and was characterized in-situ using the cesium magnetometer array by applying a well known, oscillating transverse field. In this way scalar magnetometers become sensitive to transverse components of the magnetic field. In particular, we were able to measure the depolarizing gradients such as $\partial B_z/\partial x,y$ \cite{Abel2018}. 
An algorithm minimizing these depolarizing gradients was then used to determine the currents to be applied in each of the 36 trim coils. 
This algorithm made use of the field maps to choose a set of currents so that $B_{\rm T}$ and $G_{3,0}$ were sufficiently small and fulfilled the requirements set by the associated systematic effects. 
As discussed in more detail in Ref.\,\cite{ElisePaper}, this algorithm led to an increase in decoherence time of the transverse UCN polarization to $T_2=2500$~s, and thus contributed to an increase of the statistical sensitivity. 

The last steps in initiating a new magnetic field configuration were to calibrate the cesium magnetometers and to perform a spin-echo run. The spin-echo runs assisted us in disentangling the depolarization processes, as discussed in Ref.\,\cite{Abel2018}, and also validated the magnetic field established by the optimization algorithm by directly measuring the obtainable visibility without interference from the vertical-striation effects.      

Once a magnetic-field configuration had been set, we applied various small defined vertical gradients in the range of $\pm 25$~pT/cm by using a pair of calibrated trim coils. For each configuration we took data at five different nominal gradients: $\pm 25$~pT/cm, $\pm 12$~pT/cm and $< |5|$~pT/cm. These five gradients were required to perform the crossing point analysis introduced in section~\ref{CrossPointAna}. The range was chosen to balance two contradictory requirements: it had to be large enough such that the curve $d_{\rm n}^{\rm meas}$ versus $\mathcal{R}$ showed a non-zero slope, and it had to be small enough such that the nonlinearity due to the vertical striation of ultracold neutrons under gravity could be corrected without significant sensitivity loss. In practice the gradient change was applied during the daily maintenance of the ultracold neutron source\,\cite{Anghel2018}. The schedule of this maintenance was optimized so as to increase the total number of neutrons available for the nEDM apparatus. 

\begin{figure}[hhh]
\centering
\includegraphics[width=0.9\columnwidth]{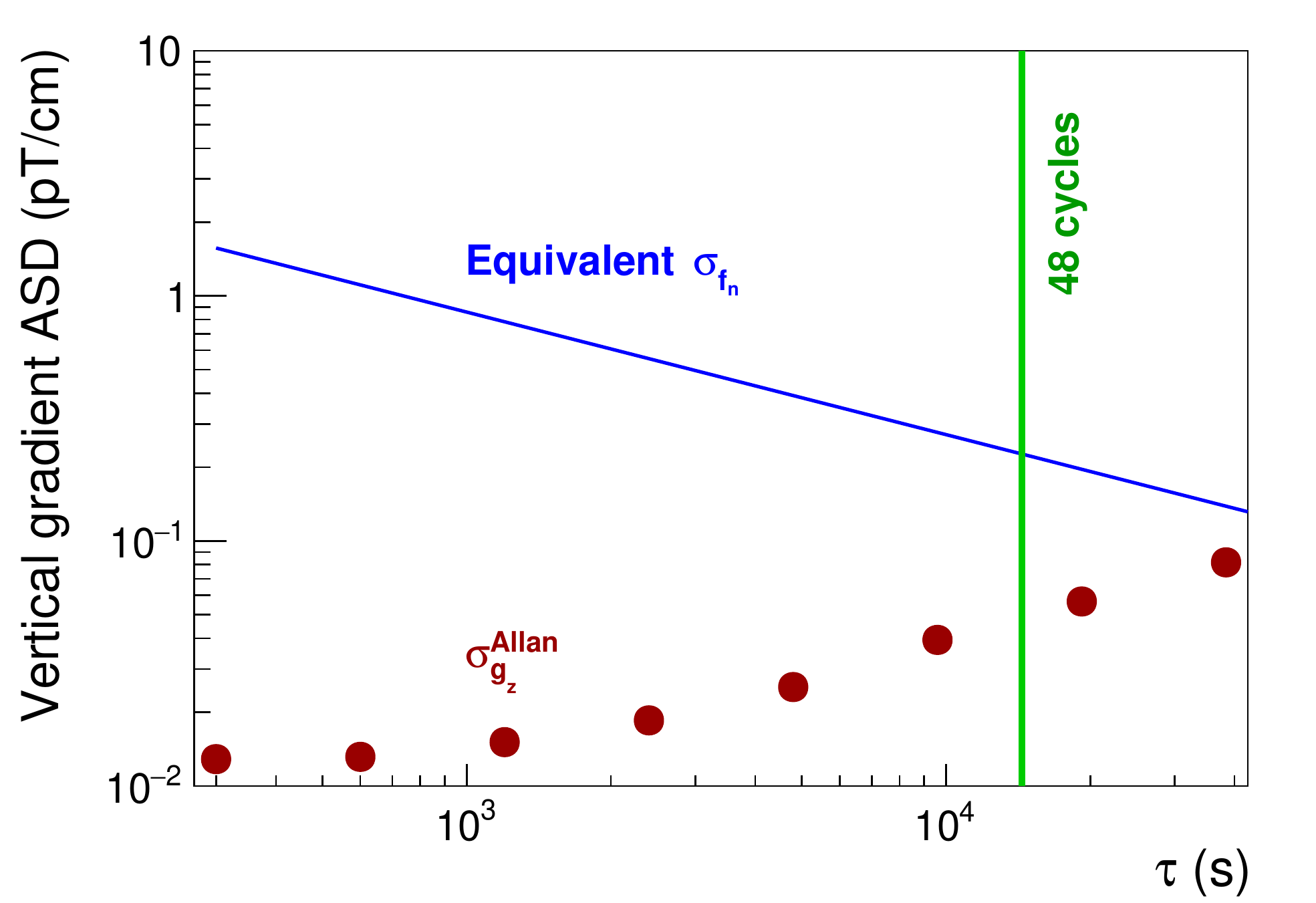}
\caption{From Ref.\,\cite{YKthesis}. The filled circles represent the Allan standard deviation of the vertical gradient for a 2.5~day long data-set. The blue line is the statistical sensitivity of the neutron precession frequency expressed in equivalent gradient fluctuations. The green vertical line shows the duration of 48 cycles (4 hours), where the fluctuation of the gradient contributes to the uncertainty at the level of a few percent of the statistical sensitivity of the neutron precession frequency. }
\label{AllanYK}       
\end{figure}

We define a data-set to be the collection of measurements with the same magnetic-field configuration and the same applied vertical gradient. Within a data-set the electric field was reversed periodically, with an optimized reversal frequency for maximum sensitivity. For our single chamber apparatus, and because we were using a mercury co-magnetometer, this frequency depended on the time stability of the vertical gradient. Indeed, fluctuations of the vertical gradient induced fluctuations of the ratio $\mathcal{R}$ (via $\delta_{\rm grav}$ in equation \eqref{Reqt}), with the potential to reduce the sensitivity. The long-term fluctuations of the vertical gradient were studied in 2014 and 2015. One can compute the Allan standard deviation of the vertical gradient $g_z$, which quantifies the fluctuations of the gradient values averaged over $\tau$:  
\begin{equation}
\sigma^{\rm Allan}_{g_z}(\tau)=\frac{1}{\sqrt{2}} \sqrt{\left < \left( g_z(t)-g_z(t+\tau) \right  )^2 \right >}.    
\end{equation}
Figure \ref{AllanYK} illustrates the status of the nEDM apparatus in early 2015 by comparing the Allan standard deviation of the vertical gradient to the statistical sensitivity of the measurement of the neutron precession frequency. 

We established in this way that a reversal frequency of 48~cycles for the electric field guarantees that the loss in sensitivity due to fluctuations of the vertical gradient will be at most a few percent. The HV reversal pattern was defined as: (+ 0 -  - 0 + ): 24+8+24+24+8+24=112 cycles. All runs started with 4 cycles at $U_{\rm HV}=0$ to initialize the online blinding of the data. The offline algorithm takes advantage of all $U_{\rm HV}=0$ cycles, to determine the central frequency of the Ramsey fringe. Furthermore, the (+ - - +) sequencing reduces the impact of a drift of  the vertical gradient. 

\begin{figure*}[h]
\centering
\includegraphics[width=13cm,clip]{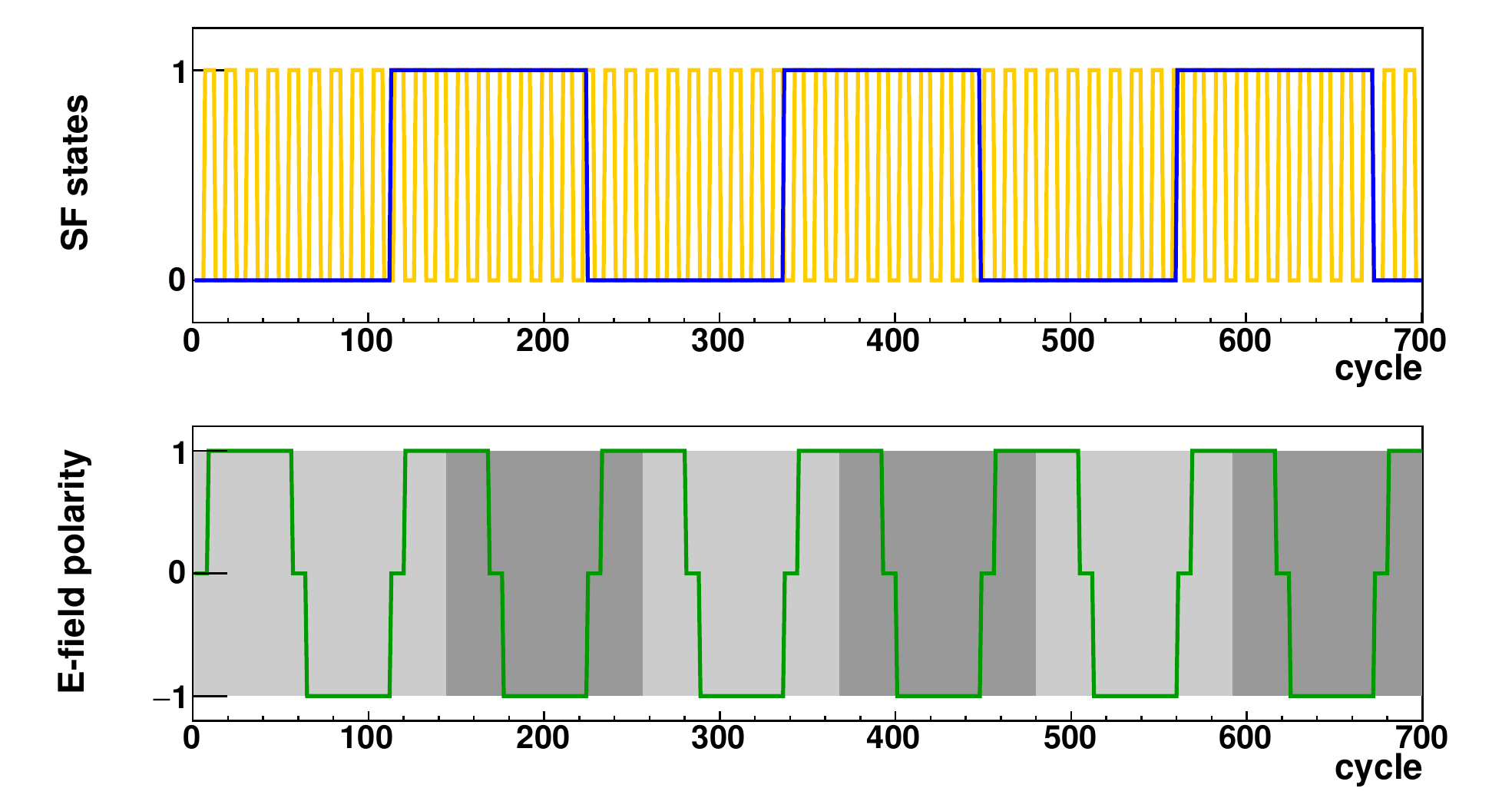}
\caption{From Ref.\,\cite{YKthesis}. State-machine representation of the nEDM data-taking sequence. The upper panel shows the state of the spin flippers as a function of the cycle number. In orange, the SF2 state changes every four cycles. In blue, the state of SF1 is seen changing every 112 cycles. In the lower panel, the HV polarity is shown in green as a function of the cycle number. The light/dark grey areas each represent a full HV pattern. }
\label{StateMachine}      
\end{figure*}

Figure~\ref{StateMachine} summarizes the sequence for the nEDM data taking. The spin flipper 2 A (B) was switched ON/OFF (OFF/ON) every 4 cycles. The high voltage was reversed every 48 cycles, and eight cycles were recorded at zero voltage at each reversal.
Hence, a full high-voltage pattern lasted for 112 cycles, a duration shorter than half a day which avoided a bias due to possible fluctuations of hidden parameters with a daily periodicity. Once a full high-voltage pattern was complete, the state of SF1 was reversed. 

\section{Sensitivity}
\label{sensitivity}
The ultimate limiting factor for the precision measurement of the neutron EDM is the accuracy with which the precession frequency of the neutron can be determined. Hence, the irreducible error on the neutron EDM is given by
\begin{equation}
\sigma(d_{\rm n}^{\rm meas})=\frac{\hbar}{2 \alpha T E \sqrt{N} \sqrt{N_{\rm cycles}}},
\label{SensitivityEqt}    
\end{equation}
where $\alpha=\alpha_0 \exp{\left(-T/T_2\right)}$ is the neutron polarization at the end of the precession time $T$, $N=N_0 \exp{\left(-T/T_0\right)}$ is the number of neutrons per cycle counted at the end of the precession time, and $N_{\rm cycles}$ is the number of cycles.
Table\,\ref{tab-sensitivity} summarizes the average values of the key parameters during data-taking at PSI, and emphasizes the improvements with respect to the previous data taking at ILL before 2006.
The increase in $\alpha$ from $\alpha(T=130)=0.58$ to $\alpha(T=180)=0.75$ is mainly due to the reduction in depolarization processes discussed above.
The largest limiting factor is the combined effect of the analyzing power of the spectrometer, characterized as $\alpha(T=0)=0.86$, and the detection efficiency.
Due to the remaining magnetic-field inhomogeneities after the optimization algorithm, the polarization after 180~s was reduced to $\alpha(T=180)=0.80$.
The most important remaining contribution to this depolarization arose from the vertical striation of ultracold neutrons under gravity, with the extent of depolarization observed depending upon the vertical gradient that was applied in conjunction with the optimized magnetic field.
By reducing the range of vertical gradients applied during data taking, the impact of this depolarizing process resulted in $0.71<\alpha(T=180)<0.80$, with an average $\alpha(T=180)=0.75$.

The apparatus was run for two years at $E=\pm 132/12$~kV/cm without any substantive problems, with a helium atmosphere at $p_{\rm He}\approx 0.5\!\times\!10^{-3}$\,mbar. 
The limiting factor for the magnitude of the electric field was the presence of the cesium-magnetometer array, which required optical fibers connecting to ground at one end and close to the high-voltage electrode at the other. Without the cesium array the apparatus was successfully operated at $E=15$~kV/cm. The improvement over time in the number of ultracold neutrons counted at the end of the precession period was mainly due to the PSI source output, which continuously increased over the years, together with an improvement of 18~$\%$ of the detector efficiency arising from the development of our simultaneous spin analyzer\,\cite{Afach2015}.

\begin{table}[ht]
\centering
\caption{Summary of the average statistical sensitivity achieved during the nEDM data taking in 2015-2016, and during the previous data taking at ILL (from \cite{Pendlebury2015}). }
\label{tab-sensitivity}       
\begin{tabular}{lr|c|c}
 && PSI 2016 & ILL 2006 \\
\hline
\hline
$\alpha$ & & 0.75 & 0.58 \\
$T$ &(s) & 180 & 130 \\
$E$ &(kV/cm) & 11 & 7\\
$N$& & 15'000 & 14'000\\
$N_{\rm cycles}$ &(cycles/day) & 288  & 400\\
\hline
$\sigma(d_{\rm n}^{\rm meas})$ &(e.cm per day) & $1.1~10^{-25}$ & $2.6~10^{-25}$ \\
\end{tabular}
\end{table}
\pagebreak
\section{Conclusion}
The data taking at the Paul Scherrer Institute using an upgraded version of the RAL-Sussex-ILL spectrometer ended in 2017. During 2015 and 2016 the spectrometer was used exclusively to measure the neutron electric dipole moment, and the total raw sensitivity achieved was better than $1\!\times\!10^{-26} e{\rm cm}$. The unprecedented suppression, measurement and control of potential systematic effects, in particular through the precise tuning of the magnetic field  and its incorporation within the data taking strategy, is the outcome of a decade of research and gives great confidence in the emerging nEDM results. 

\section*{Acknowledgements}
We acknowledge the excellent support provided by the support groups of the Paul Scherrer Institute. Dedicated technical support by M.~Meier, F.~Burri, B.~Bougard, J.~Bregeault, J.F.~Cam, B.~Carniol, P.~Desrue, D.~Etasse, C.~Fontbonne, D.~Goupilli\`ere, J.~Harang, J.~Hommet,
H.~de~Pr\'eaumont, J.~Lory, Y.~Merrer, C.~Pain, J.~Perronnel and C.~Vandamme. is gratefully acknowledged.

J.M. Pendlebury (Sussex) throughout his career made innumerable invaluable contributions to the field as a whole and to this experiment in particular.  The original nEDM spectrometer, as operated at ILL before its 2009 transport to PSI, was developed by the Sussex/RAL collaboration led by K. Smith and J.M. Pendlebury, and was funded by the UK's PPARC (now STFC).  

We gratefully acknowledge financial support from  the Swiss National Science Foundation through projects 200020-137664 (PSI), 200021-117696 (PSI), 200020-144473 (PSI), 200021-126562 (PSI), 200021-181996 (Bern), 200020-172639 (ETH) and 200020-140421 (Fribourg); and from STFC, via grants ST/M003426/1, ST/N504452/1 and ST/N000307/1. The LPC Caen and the LPSC Grenoble acknowledge the support of the French Agence Nationale de la Recherche (ANR) under reference ANR-09-BLAN-0046 and the ERC project 716651-NEDM. The Polish collaborators wish to acknowledge support from the National Science Center, Poland, under grant no. 2015/18/M/ST2/00056. P. Mohanmurthy acknowledges grant SERI-FCS 2015.0594. This work was also partly supported by the Fund for Scientific Research Flanders (FWO), and Project GOA/2010/10 of the KU Leuven. In addition we are grateful for access granted to the computing grid infrastructure PL-Grid.

\end{document}